\magnification\magstephalf
\overfullrule 0pt

\font\rfont=cmr10 at 10 true pt
\def\ref#1{$^{\hbox{\rfont {[#1]}}}$}


\font\fourteenbf=cmbx12 scaled\magstep1
\font\fourteenit=cmti12 scaled\magstep1

\font\twelvebf=cmbx12


\def\pmb#1{\setbox0=\hbox{#1}
 \kern.05em\copy0\kern-\wd0 \kern-.025em\raise.0433em\box0 }

\def\slash{/\kern-.5em}

\def \half {{\scriptstyle {1 \over 2}}}

 %


\def\boxit#1{\vbox{\hrule\hbox{\vrule\kern1pt\vbox
{\kern1pt#1\kern1pt}\kern1pt\vrule}\hrule}}

\def\h{\hfill\break}
\parskip=6pt
\parindent=0pt
\hsize=17truecm\hoffset=-5truemm
\voffset=-1truecm\vsize=24.5truecm
\def\footnoterule{\kern-3pt
\hrule width 17truecm \kern 2.6pt}


\catcode`\@=11 

\def\nolabels{\def\wrlabeL##1{}\def\eqlabeL##1{}\def\reflabeL##1{}}
\def\writelabels{\def\wrlabeL##1{\leavevmode\vadjust{\rlap{\smash%
{\line{{\escapechar=` \hfill\rlap{\sevenrm\hskip.03in\string##1}}}}}}}%
\def\eqlabeL##1{{\escapechar-1\rlap{\sevenrm\hskip.05in\string##1}}}%
\def\reflabeL##1{\noexpand\llap{\noexpand\sevenrm\string\string\string##1}}}
\nolabels
\global\newcount\refno \global\refno=1
\newwrite\rfile
\def\defref{$^{{\hbox{\rfont [\the\refno]}}}$\nref}
\def\nref#1{\xdef#1{\the\refno}\writedef{#1\leftbracket#1}%
\ifnum\refno=1\immediate\openout\rfile=refs.tmp\fi
\global\advance\refno by1\chardef\wfile=\rfile\immediate
\write\rfile{\noexpand\item{#1\ }\reflabeL{#1\hskip.31in}\pctsign}\findarg}
\def\findarg#1#{\begingroup\obeylines\newlinechar=`\^^M\pass@rg}
{\obeylines\gdef\pass@rg#1{\writ@line\relax #1^^M\hbox{}^^M}%
\gdef\writ@line#1^^M{\expandafter\toks0\expandafter{\striprel@x #1}%
\edef\next{\the\toks0}\ifx\next\em@rk\let\next=\endgroup\else\ifx\next\empty%
\else\immediate\write\wfile{\the\toks0}\fi\let\next=\writ@line\fi\next\relax}}
\def\striprel@x#1{} \def\em@rk{\hbox{}} 
\def\lref{\begingroup\obeylines\lr@f}
\def\lr@f#1#2{\gdef#1{\defref#1{#2}}\endgroup\unskip}
\newwrite\lfile
{\escapechar-1\xdef\pctsign{\string\%}\xdef\leftbracket{\string\{}
\xdef\rightbracket{\string\}}}

\def\writestop{\def\writestoppt{\immediate\write\lfile{\string\p
ageno%
\the\pageno\string\startrefs\leftbracket\the\refno\rightbracket%
\string\def\string\secsym\leftbracket\secsym\rightbracket%
\string\secno\the\secno\string\meqno\the\meqno}\immediate\closeout\lfile}}
\def\writestoppt{}\def\writedef#1{}
\catcode`\@=12 
\input epsf.tex
\rightline{DAMTP 96/66}
\rightline{M/C-TH 96/22}
\medskip
\centerline{\fourteenbf THE INTEREST OF LARGE-{\fourteenit t} ELASTIC SCATTERING}
\vskip 10mm
\centerline{A Donnachie}
\vskip 3mm
\centerline{Department of Physics and Astronomy, University of Manchester, 
Manchester M13 9PL, UK}
\vskip 6mm
\centerline{P V Landshoff}
\vskip 3mm
\centerline{DAMTP, University of Cambridge, Cambridge CB3 9EW, UK}
\vskip 10mm
{\bf{Abstract}}

Existing data for large-$t$ $pp$ elastic-scattering differential cross-sections
are energy-independent and behave as $t^{-8}$. This has been explained in 
terms of triple-gluon exchange, or alternatively through triple-singlet
exchange. A discussion is given of the problems raised by each of these 
explanations, and of the possibility that at RHIC or LHC energies the
exchange of three BFKL pomerons might result in a rapid rise with energy.
\bigskip
\bigskip

The differential cross-section for $p p$ elastic scattering at any fixed value 
of $t$ greater than 3 to 4 GeV$^2$  falls very rapidly with increasing
beam momentum, until 400 GeV/$c$. Then it flattens dramatically, and becomes
essentially energy-independent\defref\elastic{
A Donnachie and P V Landshoff, Z Physik C2 (1979) 55
}. Furthermore, its shape as a function of $t$ then becomes extremely simple,
as is seen in figure 1, where the data at five energies are plotted together
with 
$$
 d\sigma/dt = 0.09\,t^{-8}\eqno(1)
$$
(in mb GeV$^{-2}$ units).
\midinsert
\epsfxsize\hsize\epsfbox{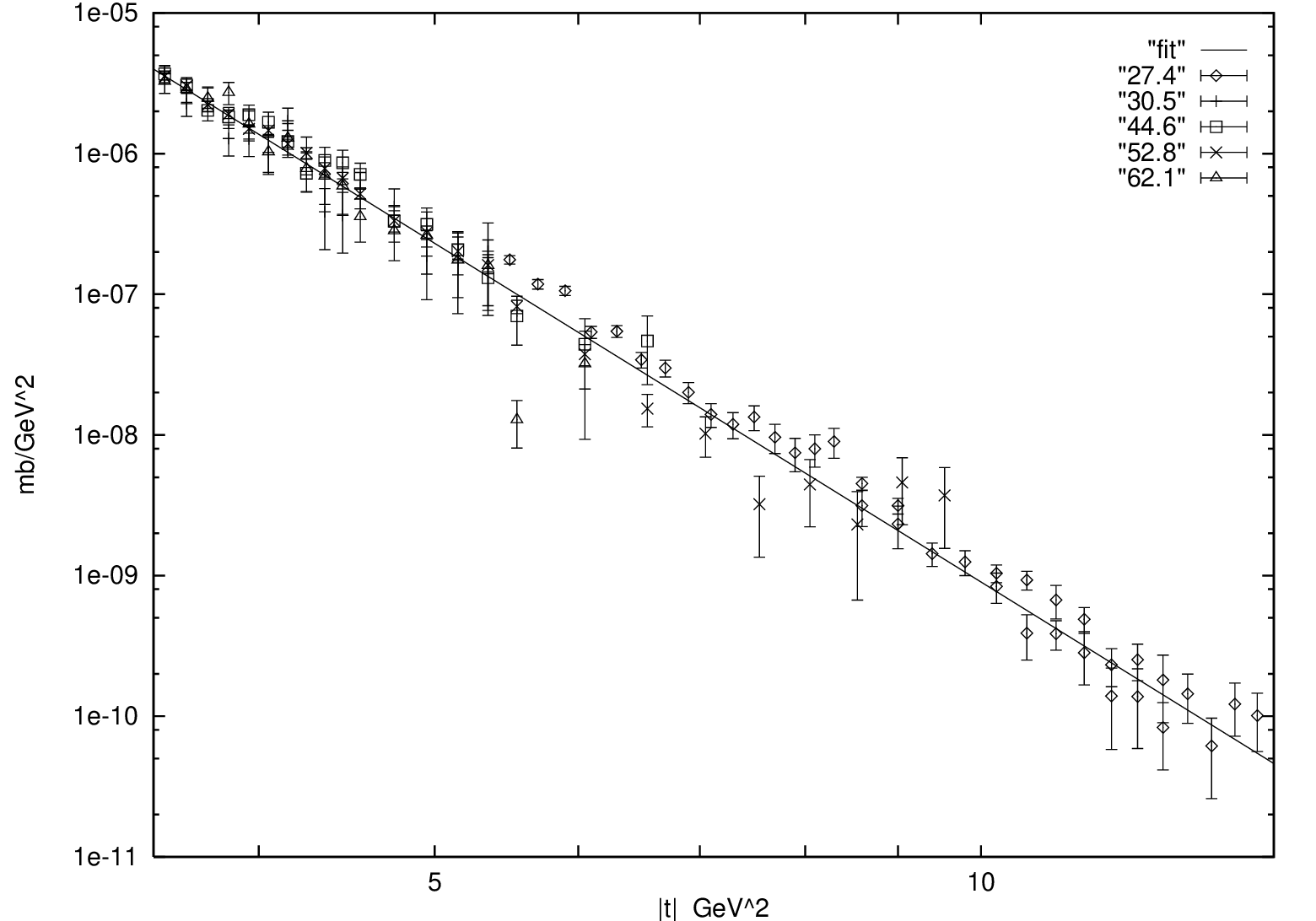}
\centerline{Figure 1: differential cross-section\defref\data{
E Nagy et al, Nuclear Physics B150 (1979) 221\h
W Faissler et al, Physical Review D23 (1981)33
} for $pp$ elastic scattering, with the fit (1)}
\vskip 1.6 truecm
\centerline{{{\epsfxsize 6truecm\epsfbox{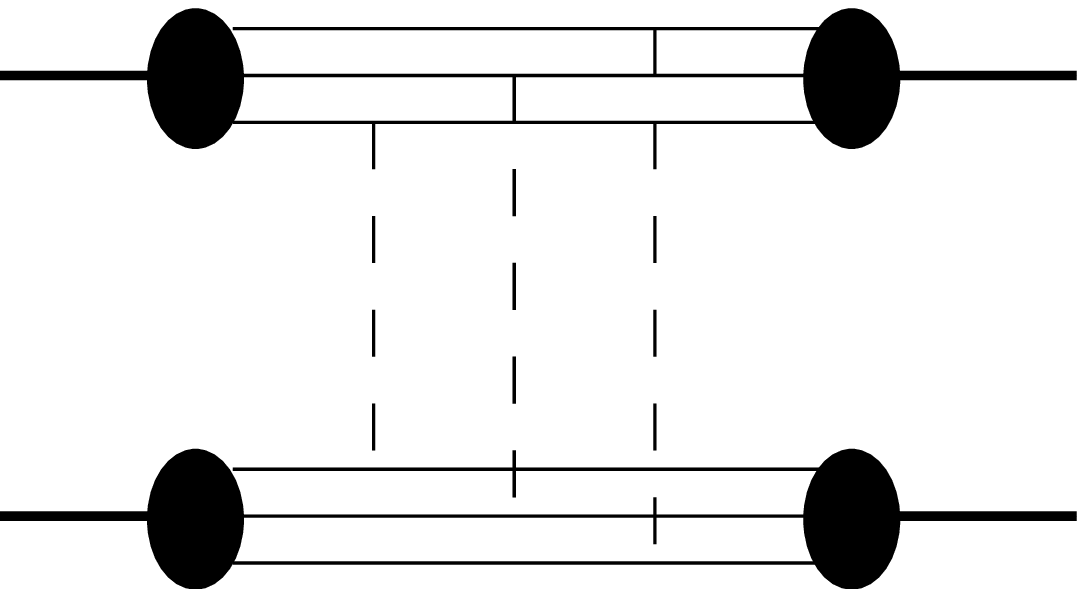}}}}
\centerline{Figure 2: lowest-order mechanism for large-$t$ elastic scattering}
\endinsert

One expects that it should be valid to apply perturbative QCD to
large-$t$ elastic scattering. In lowest order the dominant diagram is
the 3-gluon-exchange diagram of figure 2. For $s\gg|t|\gg m^2$ this
yields\ref{\elastic}
$$
 d\sigma/dt \sim \alpha_s^6\,t^{-8}
\eqno(2)
$$
One power of $t^{-2}$ arises from external kinematical factors and the
remaining $t^{-6}$ from the three gluon exchanges, each contributing 
$\alpha _s^2t^{-2}$. To obtain the result (1) it appears necessary 
to assume that 
the coupling constant $\alpha_s$ does not run. Because it occurs raised
to the sixth power, any variation with either $s$ or $t$ would cause
a problem, given the data of figure 1. 
This has long been a puzzle. Our purpose in this paper is to discuss
this, together with expectations for measurements at higher values of
$t$ and of $s$.

The energy of each incoming proton is shared among its constituent
quarks. For want of anything better, we shall assume throughout this paper
that, on average, it is shared equally. Then
the subenergy and momentum transfer associated with each quark-quark
scattering is on average
$$
\hat s\approx s/9~~~~~~~~~~~~~~~~~~\hat t\approx t/9
\eqno(3)
$$
Over the range of $t$ for the data shown in figure 1, 
this is 1.6 GeV$^2 \ge |\hat t|\ge 0.4$ GeV$^2$. 
Given this, it is clear that non-perturbative effects
should be considered. A well-motivated way to handle the non-perturbative
region has been given by Cornwall\defref\cornwall{
J M Cornwall: Physical Review D26 (1982) 1453
}, who deduced by solving 
Schwinger-Dyson equations that the contribution to quark-quark
scattering from single-gluon exchange 
can be well approximated by $\alpha _s(-\hat t)D(-\hat t)$ with
$$
D^{-1}(q^2) = q^2 + m^2(q^2)$$$$
\alpha_{s}(q^2) = {{12\pi}\over{(33-2N_{f})\log\Bigl[{{q^2+4m^2(q^2)}\over
{\Lambda^2}}\Bigr]}}\eqno(4)
$$
where the running gluon mass is given by 
$$
m^2(q^2) = m_0^2\Biggl[{{log{{q^2+4m_0^2}\over{\Lambda^2}}}\over{log{{4m_0^2}
\over{\Lambda^2}}}}\Biggr]^{-12/11}
\eqno(5)
$$
The fixed mass $m_0^2$ can be determined\defref\halzen{
A Donnachie and P V Landshoff, Nuclear Physics B311 (1988/89) 509\h
M B Gay Ducati, F Halzen and A A Natale, Physical Review D48 (1993) 2324
} from the condition that the
simple exchange of a pair of gluons between quarks is the zeroth-order
approximation to soft pomeron exchange at $t = 0$. This requires that
the integral 
$$
\beta_0^2 = {{4}\over{9}}\int d^2q[\alpha_s(q^2)D(q^2)]^2
\eqno(6)
$$
be about 4 GeV$^{-2}$. With a choice of $\Lambda = 200$ MeV this gives
$m_0 = 340$ MeV.

Over the $\hat t$-range of 
interest the variation of $\alpha_s$ as defined by (4)
and the departure of $D(-q^2)$ from $1/q^2$ work in opposite directions
and their product still varies approximately as $1/q^2$.
So this modified quark-quark scattering amplitude still provides a good
fit to the data.
The prediction
of energy independence is unaffected. At larger values of $t$
the effect of the running coupling does become apparent, but only
very very slowly: see figure 3.
\midinsert
\epsfxsize\hsize\epsfbox{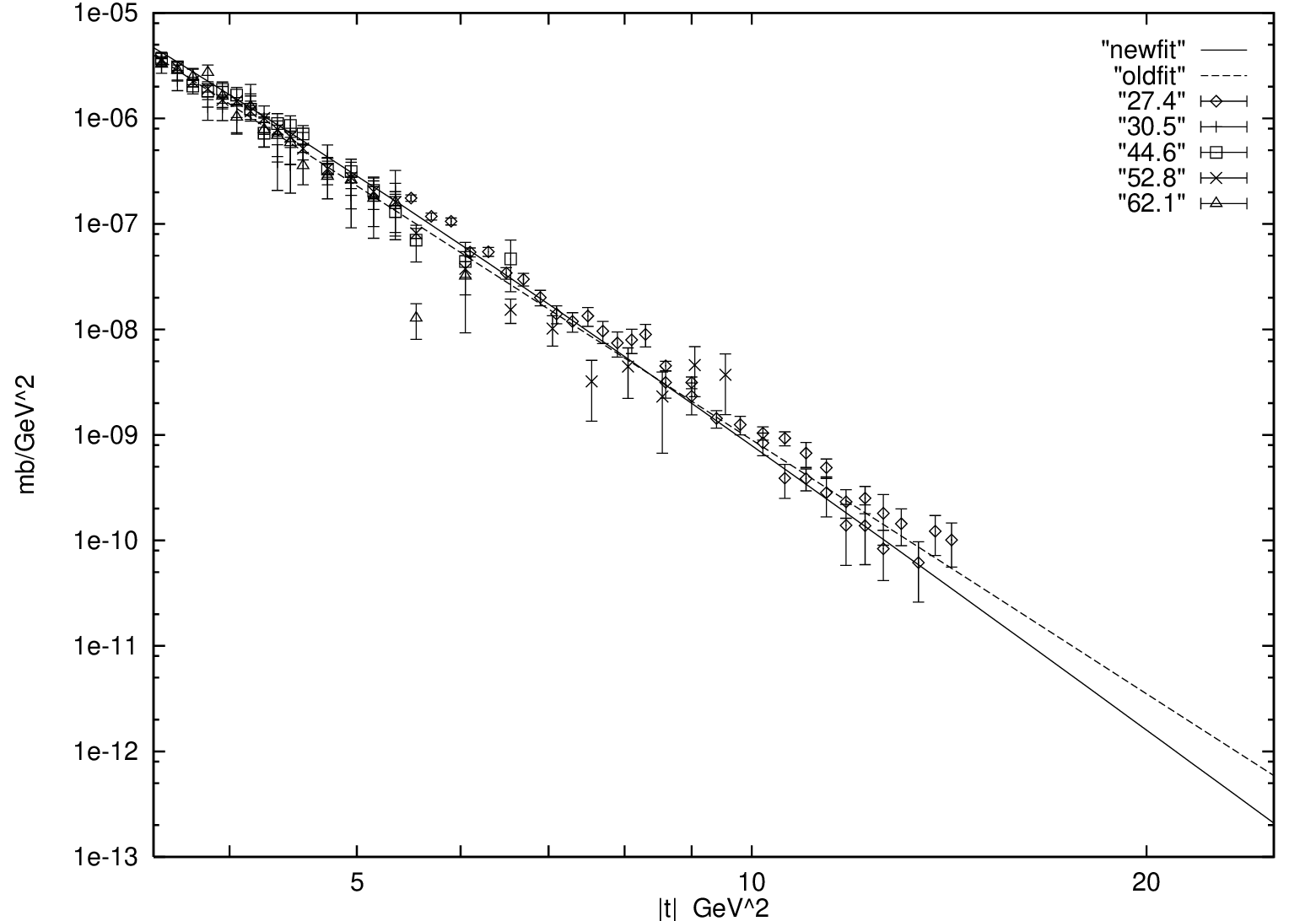}
\centerline{Figure 3: effect of including running coupling and running mass
in the fit of figure 1}
\endinsert

The data shown in figure 3 begin at $\surd{|t|}=1.9$ GeV. This momentum 
transfer is shared among the three gluons, and so the momentum transfer $\hat t$
associated with each quark-quark scattering extends down to quite small
values.
At small $\hat t$, 
high-energy quark-quark scattering is dominated by soft pomeron
exchange. The corresponding amplitude is\ref{\elastic}
$$
i\beta _0^2\gamma\cdot\gamma (\alpha '\hat se^{-\half i\pi})^{\alpha (\hat t)-1}
$$$$
\alpha(t) =1+ \epsilon_0 + \alpha' t
\eqno(7)
$$
with $\epsilon = 0.08$, $\alpha' = 0.25$ GeV$^{-2}$
and $\beta_0^2 = 4$ GeV$^{-2}$.
We must discuss whether we should include also contributions where we
replace either one of the gluons in
figure 2, or all three of them,  with a soft pomeron.
Thus the amplitude becomes
$$
A(s,t)=A_{ggg}(t)+A_{ggP}(s,t)+A_{PPP}(s,t)$$$$
A_{ggg}(t)={N\over t}{5\over 54}
\left (4\pi\alpha_s(-\hat t)D(-\hat t)\right )^3$$$$
A_{ggP}(s,t)={N\over t}{1\over 6}\left (4\pi\alpha_s(-\hat t)D(-\hat t)
\right )^2\left (i\beta _0^2(\alpha '\hat se^{-\half i\pi})^{\alpha (\hat t)-1}
\right )$$$$
A_{PPP}(s,t)={N\over t}\left (i\beta _0^2(\alpha '\hat se^{-\half i\pi})^
{\alpha (\hat t)-1}\right )^3
\eqno(8)
$$
where $N$ is a constant, fitted to the data, whose value in principle
could be calculated from knowledge of the proton wave function. We are
still making the approximation that the energy of each proton is
shared equally among its three quarks, so that $\hat s=s/9$ and
$\hat t=t/9$. We can only use the amplitudes $A_{ggP}$ and $A_{PPP}$ in
(8) for $|\hat t|$ less than about 0.7 GeV$^2$,
that is $|t|$ less than about 6 GeV$^2$,
because\ref{\elastic} for larger values
single-soft-pomeron exchange begins to become significantly reduced
by double-soft-pomeron exchange. Within this range of $t$-values,
including the terms $A_{ggP}$ and $A_{PPP}$ has very little effect,
and it decreases as the energy increases because, according to (7),
$\alpha (\hat t)-1$ is negative for $|\hat t|>0.32$ GeV$^2$. 
At the left-hand side
of figure 3 their combined contribution would be less than 10\%
to the differential cross-section. Thus our fit of figure 3,
which includes just triple-gluon exchange, is largely unaffected.

An alternative viewpoint has been put by Sotiropoulos and 
Sterman\defref\sterman{
M G Sotiropoulos and G Sterman, Nuclear Physics B419 (1994) 59
and B425 (1994) 489
},
again within the context of the triple quark-scattering model. At the 
level of quark-quark elastic scattering with multiple-gluon exchange, 
they find  an evolution in $t$ that, in leading log approximation, 
becomes diagonal in a singlet-octet basis in the $t$-channel as $s 
\rightarrow \infty$. The octet exchange in the hard scattering is 
Sudakov-suppressed with the standard reggeized $s^{\alpha_{g}(t)}$ behaviour. 
In contrast, the Sudakov suppression for 
the $t$-channel singlet exchange in the 
hard scattering is $s$-independent. This difference results in the
suppression of the octet exchange relative to the singlet exchange.

The lowest order singlet exchange is simply two-gluon exchange, that is 
a $C = +1$ exchange. Sotiropoulos and Sterman 
choose a model in which the large-momentum-transfer 
quark-quark amplitude is dominated by singlet exchange,
which they estimate to approximate to
$(\alpha_s^2/\hat t)$ within the appropriate range of $\hat t$.
There is an ambiguity in the explicit form of the
hard singlet amplitude to lowest order because it is already IR divergent
to this order. Different IR subtraction procedures yield different
expressions for the singlet hard scattering amplitude, and they can 
only be fixed by considering the amplitude at higher order in $\alpha_s$.
At lowest order it is necessary to introduce an IR cutoff, which is
arbitrary at the quark-quark scattering level but which does have
a physical meaning when the quarks are embedded in a proton and is
related to the transverse size of the proton. Thus the transverse 
structure of the proton wave function is an essential feature of the
calculation, it is intrinsically non-perturbative and introduces an
arbitrary parameter. It is possible to reproduce a behaviour close
to the $t^{-8}$ of the $p p$ data for particular choices of the
non-perturbative proton wave function, the IR cutoff parameter
and $\Lambda_{QCD}$. Note, however, that Sotiropoulos and Sterman
exchange two gluons between each pair of quarks, and so for them 
$d\sigma /dt\propto \alpha _s^{12}$, instead of $\alpha _s^{6}$ as
in (2). So any running of $\alpha _s$ would have a particularly noticable
effect.

One might argue that the simple exchange of two  perturbative gluons is
not appropriate as this should be used rather as the input to the BFKL
equation\defref\bfkl{
E A Kuraev, L N Lipatov and V Fadin, Soviet Physics JETP 45 (1977) 199\h
Y Y Balitskii and L N Lipatov, Sov J Nucl Phys 28 (1978) 822
}, which would convert the energy-independence of the two-%
gluon exchange to a strong energy dependence and invalidate the
argument for this triple $C = +1$ exchange providing the explanation
for the existing 
$p p$ data at $-t \ge 3.5$ GeV$^2$. One can also be more pragmatic
and ask whether there is evidence in the $p p$ data which allows one to
determine whether $C = -1$ or $C = +1$ exchange dominates in 
this range of $t$.

It is well established that there is an important $C = -1$ exchange in 
$p p$ and $p \bar p$ scattering at $-t \sim 1.35$ GeV$^2$, as the sharp dip 
in the differential cross section observed in ISR data for the former 
process\ref{\data} is absent in the latter\defref\breakstone{
A Breakstone et al, Physical Review Letters 54 (2180) 1985
}. It is natural to suppose that this $C=-1$ exchange survives at
larger values of $t$, and a consistent picture\defref\elastic{
A Donnachie and P V Landshoff, Nuclear Physics B267 (1986) 690
} of $p p$ and $p \bar p$ scattering at all values of $t$
can be constructed on the basis that the $C = -1$ exchange that helps
to give the $pp$ dip is just the three-gluon-exchange mechanism of
figure 2, and that the same mechanism dominates the
large-t data. Indeed, this even led us to predict the absence of a dip
in $\bar pp$ scattering\defref\predict{
A Donnachie and P V Landshoff, Physics Letters 123B (1983) 345
} before the measurements were made. However 
the data for $-t \ge 3.5 GeV^2$ alone cannot unambiguously
distinguish $C = -1$ exchange from $C = +1$ exchange.

Nevertheless, one can deduce, by considering the data at lower $t$, that the 
the Sotiropoulos-Sterman triple colour-singlet exchange is
unlikely to be an adequate replacement for our triple-gluon exchange as
the explanation for the 
existing large-$t$ elastic $p p$ data.  In order to describe the large-$t$
data, it would need to be larger than soft-pomeron exchange, in the same
way as we have argued above that gluon exchange must be if it is to provide the 
explanation. The key difference that distinguishes Sotiropoulos-Sterman exchange
from gluon exchange is that if it can occur as triple exchange 
in $pp$ elastic
scattering it must also be able to occur as single exchange. The single
exchange carries all the momentum transfer. So if in triple exchange
at $|t|=3.5$ GeV$^2$ the Sotiropoulos-Sterman exchange dominates over
soft-pomeron exchange, then the same must be true for single exchange
at $|t|\approx 0.4$ GeV$^2$. The data at such a small value of $t$
do not support this at all\ref{\elastic}. 

Nevertheless, the obvious question is whether the full BFKL pomeron\ref{\bfkl},
rather than its Sotiropoulos-Sterman truncated version, will become
apparent in large-$t$ elastic scattering at the much higher energies that
will be attainable at RHIC or the LHC.  After all, the energy $\surd\hat s$
of each quark-quark scattering is no more than about 20 GeV in the present
data, which is surely very far from asymptotic. 
The trajectory of the BFKL pomeron is surely much flatter than that of the
soft pomeron, as well as having a larger intercept. So even at
large $\hat t$ the contribution from  BFKL exchange will rise rapidly with
energy. Tevatron data place severe constraints\defref\cudell{
J R Cudell, A Donnachie and P V Landshoff, hep-ph/9602284
} on the magnitude of the BFKL-exchange contribution to the total cross-section,
that is to the amplitude at zero momentum transfer. Also, soft-pomeron
phenomenology\ref{\elastic} describes the differential cross-section well
at small nonzero momentum transfers. But this does not limit what might
be the size of the BFKL contribution at  larger momentum transfers. So,
while it is unlikely that at RHIC or LHC energies triple-BFKL exchange
is significant for $|t|$ as small as 3.5 GeV$^2$, by $|t|=10$ GeV$^2$,say,
it may well have a dramatic effect. On the other hand, the triple-gluon 
exchange, which we have argued is the explanation for the low-energy data, may
well be Sudakov-suppressed\defref\pritchard{
D J Pritchard and P V Landshoff, Z Physik C6 (1980) 69
}\ref{\sterman} at higher energies. So it could be that a dramatic rise with 
energy at large $t$ is accompanied by a fall at not-so-large $t$.

Whether or not this turns out to be true, further data are needed in order
to elucidate the mysteries posed by the existing data: why does the very
simple fit (1) work so well?

\bigskip
{\it This research is supported in part by the EU Programme ``Human Capital
and Mobility", Network ``Physics at High Energy Colliders'', contract
CHRX-CT93-0357 (DG 12 COMA), and by PPARC.}

\vfill\eject
\medskip\immediate\closeout\rfile\writestoppt
\baselineskip=14pt{{\twelvebf References}}\bigskip{\frenchspacing%
\parindent=20pt\escapechar=` \input refs.tmp\bigskip}\nonfrenchspacing
\bye